\documentclass[preprint,showpacs,prl,10pt]{revtex4}

\usepackage{amsmath}
\usepackage{amsfonts}
\usepackage{amssymb}%
\usepackage{graphicx}
\usepackage{dcolumn}
\usepackage{bm}

\usepackage[dvips]{color}

\begin{document}

\definecolor{brown}{rgb}{0,1,0}

\title{Quantum mechanics of a constrained electrically charged particle\\ in the presence of electric currents}

\author{Bj\o rn Jensen}
\affiliation{Faculty of Science and Engineering, \\ Vestfold University College, N-3103 T\o nsberg, Norway}

\author{Rossen Dandoloff}
\affiliation{Laboratoire de Physique Theorique et Modelisation, \\ Universite de Cergy-Pontoise, F-95302 Cergy-Pontoise, France}

\date{\today}


\begin{abstract}

We discuss the dynamics of a classical spinless quantum particle carrying electric charge and constrained to move on a non singular static surface in ordinary three dimensional space in the presence of arbitrary configurations of time independent electric currents.  Starting from the canonical action in the embedding space we show that a charged particle with charge $q$ couples to a term linear in $qA^3M$, where $A^3$ is the transverse component of the electromagnetic vector potential and $M$ is the mean curvature in the surface. This term cancels exactly a curvature contribution to the orbital magnetic moment of the particle. It is shown that particles, independently of the value of the charge, in addition to the known couplings to the geometry also couple to the mean curvature in the surface when a Neumann type of constraint is applied on the transverse fluctuations of the wave function. In contrast to a Dirrichlet constraint on the transverse fluctuations a Neumann type of constraint on these degrees of freedom will in general make the equations of motion non separable.  The exceptions are the equations of motion for electrically neutral particles on surfaces with constant mean curvature.  In the presence of electric currents the equation of motion of a charged particle is generally non separable independently of the coupling to the geometry and the boundary constraints.
\end{abstract}

\maketitle

\section{Introduction}

A proper understanding of quantum physics on surfaces in ordinary three dimensional space has become immediate due to the recent construction of a range of different nanostructures like sheets, tubes, cones, spheres, tori and other structures. Thin wall quantization, which was introduced in the seminal papers \cite{daCosta, daCosta2}, has emerged in this context as an important topic, albeit a challenging one. Lower dimensional nanostructures exposed to external electric and magnetic fields have recently been given increased attention. However, the theoretical framework for understanding thin wall quantization in the presence of externally applied electric and magnetic fields is presently poorly understood. In this work we seek to put this framework on a firmer footing. In the process we also extend the general framework of thin wall quantization.

This work is organized as follows.  We next introduce some differential geometry.  We then briefly review the dimensional reduction of the Schr$\ddot{\mbox{o}}$dinger equation describing an electrically {\it neutral} particle in the embedding space onto an arbitrary non singular two dimensional static surface ${\cal S}$. It is shown how the well known geometry induced effective potential \cite{daCosta} emerges.  In the same vain we then consider the dimensional reduction of the Schr$\ddot{\mbox{o}}$dinger equation minimally coupled to electromagnetism \cite{Ferrari} in the general case when electric currents are present.  We show that an anormal orbital magnetic moment appears. We argue that the resulting theory therefore must bee deemed unphysical.  We point out that the inclusion of electric currents makes the separation of the Schr$\ddot{\mbox{o}}$dinger equation minimally coupled to electromagnetism into uncoupled surface and transverse components problematic in general independently of the effects of the couplings to the geometry. We then derive the quantum theory for a charged particle on ${\cal S}$ from a canonical action.  This formulation results in a theory which is not plagued by the unphysical features of the dimensionally reduced Schr$\ddot{\mbox{o}}$dinger equation. 

The theory stemming from the variational principle has some interesting features. One central property is that an electrically charged quantum particle couples to a term linear in $qA^3M$, where $M$ is the mean curvature in the surface, $A^3$ is the transverse component of the applied electromagnetic vector potential and $q$ the electric charge. However, we show that this additional term cancels exactly the curvature correction to the orbital magnetic moment of the particle which led to unphysical behavior in the dimensionally reduced theory we considered initially. Another aspect which emerges from the variational approach, and which is independent of the value of $q$, is the importance of the boundary conditions imposed on the wave function in the process of constraining the particle to a specific surface. It follows that a Neumann type of boundary constraint exists.  This will couple any particle independently of the value of the charge to $M$.  Furthermore, considering a neutral particle with Neumann type of constraint in the transverse direction it follows that separation of the equation of motion can only be attained on surfaces with constant mean curvature. We summarize our findings in the last section and comment on previous work.

\section{Geometry and dimensional reduction}

We consider a smooth two dimensional static surface ${\cal S}$ in ordinary three dimensional space. We follow the parametrization in \cite{daCosta} and chart the three dimensional embedding space with coordinates $X^i$. We write the metric as \cite{daCosta,Burgess}
\begin{eqnarray}
ds^2&=& -dt^2+G_{ij}(X^i)dX^idX^j+(dX^3)^2=\nonumber\\
&=&-dt^2+G_{ab}(x^a)dx^adx^b+(dx^3)^2\, ,
\end{eqnarray}
where $G_{ab}(x^a)$ is the metric in the surface $\cal{S}$ defined by coordinates $x^a$. We assume that we can define a normal vector field $\vec{N}$ everywhere on ${\cal S}$. The coordinate direction $x^3$ is assumed to be along $\vec{N}$ in the immediate vicinity of ${\cal S}$. Our conventions will be such that indices at the beginning of the alphabet will refer to the coordinates in the surface $x^a$, while indices in the middle of the alphabet refer to the global coordinates $X^i$. It follows that \cite{Burgess}
\begin{eqnarray}
&&G_{ab}(X^i)=g_{ab}(x^a)-2K_{ab}(x^a)x^3+K^k\, _a(x^a)g_{km}(x^a)K^m\, _b(x^a)(x^3)^2\\
&&G (X^i)=\mbox{det}G_{ab} (X^i)=g(x^a)(1-4M(x^a)x^3+(2K(x^a)+4M(x^a)^2)(x^3)^2+...)\\
&& \sqrt{G(X^i)}\equiv \sqrt{g(x^a)}\xi (X^i)\, ,\, \xi (X^i) =1-2M(x^a)x^3+K(x^a)(x^3)^2+...
\end{eqnarray}
where $g_{ab}(x^a)$ is the induced metric in the surface, $g(x^a)=\mbox{det}g_{ab}(x^a)$ and $K^i\, _j(x^a)$ is the extrinsic curvature tensor associated with ${\cal S}$. $K=\mbox{det}K^i\, _j$, and $M(x^a)\equiv G^{ij}(X^i)K_{ij}(X^i)$ is the mean curvature in ${\cal S}$ (note that our definition of $M$ deviates from the one in \cite{daCosta} by a multiplicative factor $\frac{1}{2}$). 

Central to the thin wall quantization approach is the assumption of the presence of forces which constrain the particle to ${\cal S}$. It is assumed that these forces act  everywhere normal to ${\cal S}$.  We follow \cite{daCosta} and will assume that these forces can be derived from a potential $V_\lambda (X^3)$. $\lambda$ is a parameter which measures the strength of the potential. The Schr$\ddot{\mbox{o}}$dinger equation describing an electrically neutral particle in the embedding space within this framework is then given by (we use units such that $c=\hbar =1$)
\begin{eqnarray}
i\partial_t\psi &=&-\frac{1}{2m}G^{ij}\nabla_i(\nabla_j\psi )+V_\lambda (X^3)\, .
\end{eqnarray}
In order to derive a quantum theory in ${\cal S}$ we need to dimensionally reduce the Schr$\ddot{\mbox{o}}$dinger equation. We therefore decompose the covariant derivative in a coordinate gauge invariant manner as a sum of one part which acts along the surface and one term which acts normal to the surface
\begin{eqnarray}
\nabla_i=\nabla_{||i}+\nabla_{\bot i}\, .
\end{eqnarray}
The purely kinetic term in the Schr$\ddot{\mbox{o}}$dinger equation can then be written
\begin{eqnarray}
G^{ij}\nabla_i\nabla_j\psi \equiv (\nabla_{||}^2+\nabla_\bot ^2)\psi = (\partial^a\partial_a\psi +G^{ab}\, \Gamma^c\, _{ab}\partial_c\psi )+(\partial^3\partial_3\psi +G^{ab}\, \Gamma^3\, _{ab}\partial_3)\psi\, ,
\end{eqnarray}
where in the last relation we have used the coordinate gauge eq.(1). $\Gamma^i\, _{jk}$ represents the Christoffel symbols of the second kind. We will assume that the wave function is normalizable in three space such that the norm is given by
\begin{equation}
N=\int d^3X\sqrt{G}|\psi |^2=\int d^3x\sqrt{g} |\chi |^2\, .
\end{equation}
Probability conservation requires that $\psi (X^i) = \xi (x^i)^{-1/2}\chi (x^i)$. We use this relation to compute the kinetic term and rewrite the Schr$\ddot{\mbox{o}}$dinger equation in terms of $\chi$.  Clearly,
\begin{equation}
\lim_{x^3\rightarrow 0}\nabla_{||}^2\psi =\nabla_{||}^2\chi\, .
\end{equation}
We also find that
\begin{equation}
\lim_{x^3\rightarrow 0}\nabla_\bot^2\psi =\lim_{x^3\rightarrow 0}\frac{1}{\sqrt{G}}\partial_3(\sqrt{G}\partial^3\psi ) = \lim_{x^3\rightarrow 0} \xi^{-1}\partial_3(\xi\partial_3(\xi^{-1/2}\chi ))\equiv \partial_3^2\chi -V_0\chi\, .
\end{equation}
Using these relations we find in the limit $x^3\rightarrow 0$ that the Schr$\ddot{\mbox{o}}$dinger equation becomes
\begin{equation}
i\partial_t\chi =-\frac{1}{2m}\nabla_{||}^2\chi -\frac{1}{2m}\partial_3^2\chi +V_0\chi +V_\lambda\chi\, ,
\end{equation}
where $V_0$ is given by \cite{daCosta}
\begin{equation}
V_0=-\frac{1}{2m}((\frac{M}{2})^2-K)\, .
\end{equation}
We see that an effective potential has emerged depending on scalars characterizing the extrinsic curvature of ${\cal S}$. Let us apply this approach to the case when the particle is electrically charged.

The Schr$\ddot{\mbox{o}}$dinger equation in the embedding space describing an electrically charged particle coupled to static electric and magnetic fields in an arbitrary coordinate gauge is given by 
\begin{eqnarray}
i(\partial_t-iqA_t)\psi &=&-\frac{1}{2m}G^{ij}(\nabla_i-iqA_i)(\nabla_j-iqA_j)\psi \nonumber\\
&=&-\frac{1}{2m}(\nabla^i\nabla_i\psi -iqG^{ij}(\nabla_iA_j)\psi -2iqA^j\nabla_j\psi -q^2A^iA_i\psi )+V_\lambda\psi\, .
\end{eqnarray}
The wave function $\psi$ and the electromagnetic vector potential are functions of $X^i$. The scalar potential $V$ is defined as usual by $V=-A_t$.  Equation (13) is invariant under a $U(1)$ gauge-transformation of the vector potential combined with a transformation of the wave-function
\begin{equation}
\left\{
\begin{array}{l}
A_m(X^i)\rightarrow A_m'(X^i)=A_m(X^i)+\nabla_mf(X^i)\\
\psi (X^i)\rightarrow \psi '(X^i)=\psi (X^i) e^{iqf(X^i)}
\end{array}\right.
\end{equation}
where $f$ is a sufficiently well behaved scalar function. We follow the decomposition scheme in eq.(6) and decompose the electromagnetic potential in one term $\vec{A}_{||}$ which is living in the surface ${\cal S}$, and one term which is normal to ${\cal S}$, $\vec{A}_{\bot}$
\begin{equation}
\vec{A}=\vec{A}_{||}+\vec{A}_{\bot}=\vec{A}_{||}+|\vec{A}_{\bot}|\vec{N}\, .
\end{equation}
We can then write
\begin{eqnarray}
G^{ij}(\nabla_iA_j)=\nabla_iA^j=\nabla_iA^i_{||}+(\nabla_i|\vec{A}_\bot|)N^j+|\vec{A}_\bot |G^{ij}\nabla_iN_j\, .
\end{eqnarray}
All quantities are at this level expressed in terms of the global coordinates, but note that we have not specified the coordinate gauge.  
We then find that
\begin{eqnarray}
G^{ij}(\nabla_iA_j)=\nabla_iA^i=\nabla_iA^i_{||}+(\nabla_i|\vec{A}_\bot|)N^i+|\vec{A}_\bot| G^{ij}(\nabla_{||i}N_j+\nabla_{\bot i}N_j).
\end{eqnarray}
The last two terms in eq.(17) can be computed explicitly since $\nabla_{||i}N_j=K_{ij}$,
and $\nabla_{\bot i}N_j=0$ by construction. All quantities are evaluated on ${\cal S}$. Hence, the electromagnetic vector potential and thus the charged particle couple to the extrinsic curvature tensor defined on ${\cal S}$. We now partially gauge fix the coordinates, and identify the normal vector field with the unit tangent vector field of appropriately chosen coordinate lines $\vec{N}=\vec{e}_3$ such that $|\vec{A}_\bot |=A^3$. On $\cal S$ we then get the following expression for the divergence term in eq.(17)
\begin{eqnarray}
\nabla_iA^i=\nabla_{||}\cdot\vec{A}_{||}+\partial_3A^3+2A^3M\, .
\end{eqnarray}
The deduction above can also be made using the parametrization eq.(1) from the outset. In that coordinate gauge the divergence term can be written in the following way
\begin{eqnarray}
G^{ij}\nabla_iA_j&=G^{ab}(\partial_aA_b+\, \Gamma^c\, _{ab}A_c)+(\partial_3A_3+G^{ab}\, \Gamma^3\, _{ab}A_3)\, .
\end{eqnarray}
In the limit that the particle is constrained to the surface we thus have
\begin{eqnarray}
\lim_{x^3\rightarrow 0} G^{ij}\nabla_iA_j = g^{ab}\nabla_{||a}A_{||b}+\partial_3A^3+2A^3M
\end{eqnarray}
in complete agreement with the coordinate gauge invariant derivation. Here we have used 
\begin{eqnarray}
G^{33}G^{ab}\, ^{\bf{3}}\Gamma^3\, _{ab}=-\frac{1}{2}G^{ab}G_{ab,3}=-\frac{1}{2}G^{ab}(-2K_{ab}+...)=G^{ab}K_{ab}=2M\, ,
\end{eqnarray}
and
\begin{eqnarray}
&& \Gamma^3\, _{33}= \Gamma^d\, _{33}=0\, ,\, \Gamma^3\, _{ab}=-\frac{1}{2}G^{33}G_{ab,3}\, .
\end{eqnarray}
We have also assumed that $x^3\rightarrow 0$ in the last two expressions. We now express the Schr$\ddot{\mbox{o}}$dinger equation in terms of $\chi$. After performing all derivate operations and letting $x^3\rightarrow 0$ the following effective potential emerges
\begin{equation}
V_{\mbox{\small Eff}}=V_{0}-\frac{q^2}{2m}\vec{A}^2-\frac{qi}{2m}(\nabla_{||}\cdot\vec{A}_{||}+\partial_3A^3+2A^3M)\, .
\end{equation}
$V_{0}$ is the effective potential in eq.(12). In addition to the usual second order kinetic terms in the Schr$\ddot{\mbox{o}}$dinger equation in the limit $x^3\rightarrow 0$ we also get the following first order terms when the $A^j\nabla_j\psi$ term in eq.(13) is expanded 
\begin{equation}
-\frac{qi}{m}(\vec{A}_{||}\cdot\nabla_{||}+A^3(M+\partial_3))\chi =-\frac{qi}{m}(\vec{A}\cdot\nabla + A^3M)\chi\, .
\end{equation}
The linear coupling to $M$ stems from the $\xi$-factor in the factorization of the wave function. This coupling appears as a curvature contribution to the orbital magnetic moment of the particle. We now make a brief detour to discuss the externally applied field.

A natural question to pose is whether it is possible to remove the $A^3$ component of the gauge potential altogether so as to remove the coupling to $M$ completely from the Schr$\ddot{\mbox{o}}$dinger equation \cite{Ferrari}. It is apparently possible to perform a restricted gauge transformation in the Lorentz gauge such that $A^{\prime 3}=0$ \cite{Ferrari} (we note that the corresponding gauge choice in \cite{Ferrari} is not necessarily a {\it restricted} gauge transformation in the Lorentz gauge, however). Clearly, when doing this one implicitly neglects the sources of the electromagnetic field.  In the stationary situation in the Lorentz gauge the equation of motion for the electromagnetic vector potential coupled to the sources $\vec{J}$ reduces to
\begin{equation}
\nabla^2\vec{A'}\sim -\vec{J}\, .
\end{equation}
$\vec{J}$ is the effective three dimensional electric current. Trivially, setting $A^{\prime 3}=0$ will imply a vanishing source component in the direction perpendicular to the surface. With $J^3=0$ the Lorentz gauge condition implies $\partial_3A^{\prime 3}=0$ as an identity.  It is straightforward to see this.  Since we always can find a function $f(X^i)$ such that $A^{\prime 3}\rightarrow A^{\prime 3}+\partial^3f=0$ we immediately get that $\partial_3A^{\prime 3}$ transforms as $\partial_3A^{\prime 3}\rightarrow\partial_3A^{\prime 3} +\partial_3\partial^3f=\partial_3A^{\prime 3} -\partial_3A^{\prime 3}=0$. Contrary to the reasoning in \cite{Ferrari} there is no need to make any assumptions on the form of $A^3$ in order to make the normal derivate term vanish when working consistently in the Lorentz gauge. Setting $A^{\prime 3}=0$ makes the coupling to the mean curvature vanish and we regain the results in \cite{Ferrari}.  In this situation the Schr$\ddot{\mbox{o}}$dinger equation is trivially separable into a surface component and a component perpendicular to the surface \cite{Ferrari}. However, in the general situation we do have source currents in all space directions. This implies in particular that the coupling to the mean curvature will in general be present and none of the components of the vector potential can be set identically to zero.  Since the $A^{\prime 3}$-component depends on both the coordinates in the surface as well as on $x^3$ it follows trivially that the Schr$\ddot{\mbox{o}}$dinger equation is not separable in general due to the $\partial_3$ term in eq.(24).

When imposing a gauge condition on the electromagnetic vector potential it is natural to impose the Lorentz gauge condition since it is coordinate independent.  Hence, we transform to new electromagnetic gauge potentials $A^{\prime i}$ such that
\begin{equation}
\nabla\cdot\vec{A'} =\nabla_{||}\cdot\vec{A'}_{||}+\partial_3A^{\prime 3}+2A^{\prime 3}M=0\, .
\end{equation}
This condition seemingly removes the explicit coupling to the mean curvature in the effective potential.  However, now the effective potential in the surface is implicitly coupled to $M$ due to the presence of the $\vec{A}^2$ term in the potential since we can always in principle re-express one of the components of the vector potential in terms of the other two components and the mean curvature. 

When the charge is constrained to an arbitrary surface we would intuitively expect that the momentum perpendicular to the surface approaches zero due to the squeezing potential ($\partial_3\chi\rightarrow 0$, i.e.), at least in the classical limit.  However, we see that the orbital angular moment nevertheless receives a curvature correction since $A^3\neq 0$ on ${\cal S}$ in general.  The exceptions are when the particle is constrained to a minimal surface ($M=0$). On such surfaces the orbital magnetic moment of the charge is apparently independent of $A^3$.  These results are surprising and has to our knowledge not been observed in any experiment.  We interpret this result as signaling a possible breakdown of the thin wall approach to quantum theory on surfaces when dealing with electrically charged particles. We therefore next turn to a derivation of a framework for thin wall quantization when the particle is minimally coupled to electromagnetism.
 
\section{A general framework}

In this section we re-derive the quantum theory in ${\cal S}$ from a variational principle. We assume the canonical action in the embedding space (neglecting $V_\lambda$, for simplicity)
\begin{eqnarray}
S&=&\int_\Omega (-\psi^*iD_t\psi +\frac{1}{2m}(D_k\psi )^*D^k\psi ) =\\
&=&\int_\Omega (-\psi^*iD_t\psi +\frac{1}{2m}\nabla_k\psi^*\nabla^k\psi +\frac{q^2}{2m}\vec{A}^2\psi^*\psi +\frac{1}{2m}A^k C_k)\, ,
\end{eqnarray}
where we have defined
\begin{equation}
D_k\equiv\nabla_k-iqA_k\, ,\, D_t\equiv\partial_t-iqA_t\, ,\, C_k\equiv iq(\psi^*\nabla_k\psi -(\nabla_k\psi )^*\psi )\, .
\end{equation}
The integration measure is the canonical one and will be suppressed in the following. $\Omega$ is the region of integration. The electromagnetic vector potential will be treated as an external classical background. The action can be rewritten so as to exhibit the Schr$\ddot{\mbox{o}}$dinger equation for $\psi$ by utilizing
\begin{eqnarray}
\int_\Omega A^kC_k&=&iq\int_\Omega (A^k\psi^*(\nabla_k\psi )-iq\int_\Omega (\nabla_k(A^k\psi^*\psi )+(\nabla_kA^k)\psi^*\psi +A^k(\nabla_k\psi )\psi^*) =\nonumber\\ &=& iq\int_\Omega (-\nabla_k(A^k\psi^*\psi )+(\nabla_kA^k)\psi^*\psi +2A^k(\nabla_k\psi )\psi^*)\, ,
\end{eqnarray}
so that the action reads
\begin{eqnarray}
S&=&\int_\Omega (-\psi^*iD_t\psi -\frac{1}{2m}\psi^*\nabla_k\nabla^k\psi +\frac{q^2}{2m}\vec{A}^2\psi^*\psi +\frac{iq}{m}\psi^*A^k\nabla_k\psi +\frac{iq}{2m}(\nabla_kA^k)\psi^*\psi -\frac{iq}{2m}\nabla_k(A^k\psi^*\psi ))+\nonumber\\
&+&\frac{1}{2m}\int_{\partial\Omega}\psi^*\nabla^k\psi\, .
\end{eqnarray}
We have also performed a partial integration so as to isolate the dynamical term. The fourth term in ${\cal S}$ will give rise to the anormal orbital magnetic moment we noted previously.  We note that the last term in the $\Omega$-integral is absent in the canonical Schr$\ddot{\mbox{o}}$dinger equation which we used earlier.  This term contains a $\xi$-factor such that we get
\begin{equation}
\lim_{x^3\rightarrow 0}\int_\Omega\nabla_k(A^k\psi^*\psi )=2\int_\Omega A^3M\chi^*\chi +\lim_{x^3\rightarrow 0}\int_{\partial\Omega}A^k\psi^*\psi\, .
\end{equation}
Hence, this term contributes besides a surface integral, as expected, also a volume integral which adds the term
\begin{equation}
-\frac{iq}{m}A^3M\chi^*\chi\, 
\end{equation}
to the volume part of the action. Interestingly, this addition to the surface term cancels exactly the anormal orbital magnetic moment to the particle which emerges from the $A^3\nabla_3\psi$-term. Hence, in the limit $x^3\rightarrow 0$ we get
\begin{eqnarray}
S&=&\int_\Omega (-\chi^*iD_t\chi -\frac{1}{2m}\chi^*\nabla_k\nabla^k\chi +V_0\chi^*\chi+\frac{q^2}{2m}\vec{A}^2\chi^*\chi +\frac{iq}{2m}\chi^*A^k\nabla_k\chi+
\frac{iq}{2m}(\nabla_kA^k)\chi^*\chi) +\nonumber\\
&+&\lim_{x^3\rightarrow 0}\frac{1}{2m}\int_{\partial\Omega}\psi^*D^k\psi\, .
\end{eqnarray}
Clearly, eq.(30) treats $\psi$ and $\psi^*$ asymmetrically.  We could of course also have written 
\begin{eqnarray}
\int_\Omega A^kC_k&=&iq\int_\Omega (\nabla_k(A^k\psi^*\psi )-(\nabla_kA^k)\psi^*\psi -A^k(\nabla_k\psi )\psi^* -iq\int_\Omega A^k\psi^*(\nabla_k\psi )) =\nonumber\\ &=& iq\int_\Omega (\nabla_k(A^k\psi^*\psi )-(\nabla_kA^k)\psi^*\psi -2A^k(\nabla_k\psi )\psi^*)\, .
\end{eqnarray}
This form is suitable to use in the action for $\psi^*$, $S^*$. Using this $S^*$ can (in the limit $x^3\rightarrow 0$) be written as
\begin{eqnarray}
S^*&=&\int_\Omega (+\chi iD_t\chi^* -\frac{1}{2m}\chi\nabla_k\nabla^k\chi^* +V_0\chi^*\chi+\frac{q^2}{2m}\vec{A}^2\chi^*\chi -\frac{iq}{2m}\chi^*A^k\nabla_k\chi-
\frac{iq}{2m}(\nabla_kA^k)\chi^*\chi) +\nonumber\\
&+&\lim_{x^3\rightarrow 0}\frac{1}{2m}\int_{\partial\Omega}\psi(D^k\psi )^*\, .
\end{eqnarray}
Varying $S$ gives the following dynamical equation for $\psi$ in an arbitrary electromagnetic gauge 
\begin{equation}
-iD_t\chi -\frac{1}{2m}\nabla_k\nabla^k\chi +V_0\chi +\frac{q^2}{2m}\vec{A}^2\chi +\frac{iq}{2m}A^k\nabla_k\chi+
\frac{iq}{2m}(\nabla_kA^k)\chi=0\, .
\end{equation}
A similar equation for $\psi^*$ follows from $S^*$. These equations are free from the pathology found in the previous section. The last term in eq.(37) (and similarly for the corresponding conjugate equation) vanishes in the Lorentz gauge, of course.

It is a priori not completely clear how one should treat the surface integrals.  Let us for simplicity first consider closed surfaces. Since the scheme is to squeeze the particle such that it skims ${\cal S}$ it is natural to consider regions of integration $\Omega$ which are symmetrical about ${\cal S}$. It is now at least two ways to treat variations in $\psi$, $\delta\psi$ (and likewise for the conjugate theory) in the limit $x^3\rightarrow 0$ normal to the surface.  One can either demand that $\delta\psi =0$ in the transverse direction (a Dirrichlet condition, i.e.), or one can allow for certain variations along the normal direction of the surface. If one assumes $\delta\psi =0$ it would correspond to a complete freezing out of any dynamics perpendicular to the surface in the limit $x^3\rightarrow 0$.  In that case the surface integral will vanish by construction.  In \cite{daCosta} it was assumed that $V_\lambda$ has the form of an infinite potential well such that $\psi$ is forced to zero a certain distance from ${\cal S}$.  This corresponds to our Dirrichlet condition. However, physically such a constraint is probably to severe and breaks with natural limits set by the uncertainty principle e.g., since it is assumed that $x^3\rightarrow 0$. Hence, on physical grounds one can argue that it is more physically consistent to assume certain variations $\delta\psi$ on $\partial\Omega$. In this case the surface integral vanishes provided 
\begin{equation}
\lim_{x^3\rightarrow 0}D_3\psi =0\, .
\end{equation}
A similar expression holds for the conjugate field. It is interesting that this last constraint is not identical to the vanishing of the generalized momentum in the direction perpendicular to the surface since the action of the gradient operator on $\psi$ will generate a term proportional to the mean curvature. Hence, on ${\cal S}$, and including the corresponding expression for $\psi^*$, we get
\begin{equation}
\left\{
\begin{array}{l}
(\nabla_3-2iqA_3+2M)\chi=0\, ,\\
(\nabla_3+2iqA_3+2M)\chi^*=0\, .
\end{array}\right.
\end{equation}
These constraints are valid for all surfaces ${\cal S}$, and they correspond to Neumann type of boundary conditions. They have to our knowledge not been addressed previously in the extensive literature on the dynamics of constrained quantum particles. On surfaces which are not closed we will get further constraints which depend on the boundary conditions in the directions defined by ${\cal S}$. 

We saw earlier that the Lorentz condition couples the electromagnetic vector potential to the mean curvature in ${\cal S}$. Interestingly, the surface constraints eq.(39) induce a coupling between the mean curvature, the external field and the wave function independently of the gauge choice on the electromagnetic vector potential. Furthermore, note that the Neumann type of surface constraints also hold in the limit $q\rightarrow 0$, when $D_3\rightarrow \nabla_3$. It follows immediately that the Schr$\ddot{\mbox{o}}$dinger equation is separable in the neutral case only when the mean curvature is constant. Hence, these constraints make the separation of the wave equation into surface and perpendicular components highly problematic in the general situation independently of whether the particle carries an electric charge or not.  

\section{Discussion}

In this paper we briefly review, explore and extend previous work on the quantum theory of particles constrained to surfaces in ordinary three dimensional space.  Much previous work have relied on adapting the Schr$\ddot{\mbox{o}}$dinger equation directly to curved spaces.  We explored this approach in the first part of this paper.  We found that the resulting quantum mechanics of electrically charged particles on surfaces exhibits unexpected features.  We then rederived the quantum theory on surfaces starting from the canonical Schr$\ddot{\mbox{o}}$dinger action.  In this formulation the special features in the 'straightforward' dimensional reduction approach are absent. 

Part of any variational approach is the appearance of boundary terms.  In the seminal papers \cite{daCosta,daCosta2} these were not explicitly discussed, nor (to our knowledge) in the by now extensive literature on thin wall quantization.  The form of the constraining potential $V_\lambda$ employed in \cite{daCosta} corresponds to imposing the Dirrichlet condition on the transverse fluctuations of the wave function.  In the case of a neutral particle our variational approach and the one in \cite{daCosta} both imply that the wave function is always separable when the Dirrichlet condition is applied. However, we have shown that imposing the Neumann type of constraint on the transverse fluctuations of the wave function of a neutral particle implies separability only when ${\cal S}$ has constant mean curvature. This means in particular that basic geometric shapes like spheres, cones and cylinders as well as all minimal surfaces like the catenoid and the helicoidal surface allow separation into surface and transverse degrees of freedom independently of the particular applied boundary constraint when the particle is chargeless. 

In \cite{Ferrari} it was argued that (with reference to electromagnetic fields) {\it We find that there is no coupling between the fields and the surface curvature and that with a proper choice of the gauge, the surface and transverse dynamics are exactly separable.} In view of our more general treatment we have shown that both of these claims are generally incorrect.  The reasons for the differences between our approach and the one in \cite{Ferrari} is partially their use of the Schr$\ddot{\mbox{o}}$dinger equation in the embedding space, partially the way we decompose the covariant derivative and the vector potential into surface and perpendicular operators and partly due to our inclusion of currents. In \cite{Ferrari} none of these aspects was discussed. We have shown that besides the couplings to $M^2$ and $K$ the equation of motion of an electrically charged quantum particle also couples to the mean curvature $M$ through the $qA^3M$ term. 

The coupling to $M$ in the Schr$\ddot{\mbox{o}}$dinger equation appeared when it was written in an arbitrary gauge through the dimensional reduction of the $\nabla\cdot\vec{A}$ term.  Even though this term can be removed entirely from the equation by choosing the Lorentz gauge it will reappear in the form of a subsidiary condition and thus be implicitly present in the final form of the Schr$\ddot{\mbox{o}}$dinger equation through the $\vec{A}^2$ term. {\it Consequently, the Schr$\ddot{\mbox{o}}$dinger equation minimally coupled to electromagnetism will couple to $M$ in addition to the other two geometric scalars independently of the choice of the gauge on the electromagnetic vector potential}. We note that this coupling to $M$ comes in the form of a product $qA^3M$.  When $q\rightarrow 0$ the coupling disappears. We interpret this as it is the particle that interacts through the electric charge with an effective transverse component of the vector potential which equals $A^3M$. Hence, it is probably most consistent to interpret this as $\vec{A}$ interacting with the geometry through $M$ since the particle couples to $\vec{A}$ through $q$. Interestingly, in \cite{Encinosa} a similar coupling term was also found.  The derivation of the coupling term in that work was based on the $A^3\nabla_3\psi$ term in our eq.(1) when a number of further conditions and approximations were assumed on both the wave function and the vector potential. We have uncovered that this type of coupling appears generically and as exact results in thin wall quantum theory.

The discussion of the boundary term in connection with a neutral particle in the previous paragraphs carry over to the theory of electrically charged particles only to a certain extent.  The Dirrichlet condition does not by itself couple surface and transverse degrees of freedom of a charged particle, but the Schr$\ddot{\mbox{o}}$dinger equation does in general when the sources of the external field is included through a $A^3\partial_3$ term.  It is only when the sources are neglected it becomes possible to gauge away this coupling \cite{Ferrari}, and the coupling to the geometry through the $qA^3M$ term.  However, this conclusion does not hold if we apply the Neumann constraint. If we then also apply the Lorentz gauge, have no sources  and if we further perform a restricted gauge transformation such that $A^3=0$ in the Lorentz gauge the electrically charged particle will still couple explicitly to the mean curvature through the constraints eq.(39) in exactly the same way as an uncharged particle. {\it This implies that with the Neumann constraint applied the equation of motion of a charged particle is separable only on surfaces with constant mean curvature when no sources are present}.

Our work opens up some interesting avenues for further work. Most, if not all, previous work in the field of thin wall quantization have assumed either neutral particles effectively constrained by the Dirrichlet constraint, or electrically charged particles with the same constraint and no sources. Our work has thus uncovered that the picture provided thus far by the thin wall quantization approach of the quantum physics in surfaces may only represent a small glimpse of the actual quantum physics on surfaces.  Including sources for the vector potential will by itself give rise to a number of configurations which may contain new features.  So will the study of the Neumann constraint applied to concrete structures. One starting point for further work is to reconsider the quantum dynamics on basic nanostructures such as the sphere and the cylinder.

\end{document}